\documentclass{sf2a-conf2017}
\usepackage{graphicx}
\usepackage{hyperref}
\usepackage[]{natbib}  
\usepackage{epstopdf}
\usepackage{color}

\def\BibTeX{{\rm B\kern-.05em{\sc i\kern-.025em b}\kern-.08em
    T\kern-.1667em\lower.7ex\hbox{E}\kern-.125emX}}
\bibpunct{(}{)}{;}{a}{}{,}  %%%%%%%%%%%%%  A&A bibliography style
\begin{document}

\TitreGlobal{SF2A 2017}

\title{WebPlotDigitizer, a polyvalent and free software to extract spectra from old astronomical publications: \\
application to ultraviolet spectropolarimetry}
\runningtitle{WebPlotDigitizer for astrophysics}

\author{F.~Marin}\address{Universit\'e de Strasbourg, CNRS, Observatoire astronomique de Strasbourg, UMR 7550, F-67000 Strasbourg, France }
\author{A.~Rohatgi}\address{ANSYS Inc., Austin, Texas, USA}
\author{S.~Charlot}\address{Sorbonne Universit\'es, UPMC-CNRS, UMR7095, Institut d'Astrophysique de Paris, F-75014 Paris, France}

\setcounter{page}{237}

\maketitle

\begin{abstract}
In this contribution, we present WebPlotDigitizer, a polyvalent and free software developed to facilitate easy 
and accurate data extraction from a variety of plot types. We describe the numerous features of this numerical 
tool and present its relevance when applied to astrophysical archival research. We exploit WebPlotDigitizer to 
extract ultraviolet spectropolarimetric spectra from old publications that used the Hubble Space Telescope, Lick 
Observatory 3~m Shane telescope and Astro-2 mission to observe the Seyfert-2 AGN NGC~1068. By doing so, we 
compile all the existing ultraviolet polarimetric data on NGC~1068 to prepare the ground for further investigations
with the future high-resolution spectropolarimeter POLLUX on-board of the proposed Large UV/Optical/Infrared 
Surveyor (LUVOIR) NASA mission.
\end{abstract}


%%-----------------------------------------------------------------

\section{Introduction}
Astronomical spectra acquired before the era of Internet and on-line data storage are only available through 
papers published in physical form. Observational data, in particular those obtained by small missions and 
modest telescopes before the 90's, are not all stored in databases and several observations cannot be retrieved.
New observations are thus required but it is a time-consuming task, especially for those who only want to fit 
their numerical simulations in order to test the relevance of their model. In this context, asking for observational 
time on new satellites and telescopes that are already facing high pressure-factors is a waste of effort. 
The best solution relies on digitalization of the old spectra using high resolution scanners. The digitalized 
plots are thus available for data extraction using modern softwares. However, many of the software tools dedicated 
to this task are either expensive or not very versatile. The common complaints about the existing tools are their 
limited features, poor compatibility with different operating systems and closed source code.

In this contribution, we present WebPlotDigitizer, a software created by Ankit Rohatgi that is distributed free 
of charge as an open source web-based tool. Thanks to its large adaptability, we show that the software can be 
easily handled to accurately extract data from any published paper in the field of astrophysics (and beyond). 
To show the great potential of this tool, we use it to retrieve the old ultraviolet polarimetric data of NGC~1068, 
a Seyfert-2 active galactic nuclei (AGN), acquired before the millennium by the Hubble Space Telescope, Lick 
Observatory 3~m Shane telescope and Astro-2 mission. This preliminary step will allow us to test model predictions
for the high-resolution spectropolarimeter POLLUX that is proposed for the NASA's mission study LUVOIR.

\section{Description and application of WebPlotDigitizer}
\subsection{An overview of the software}

\begin{figure}[ht!]
 \centering
 \includegraphics[trim = 0mm 0mm 0mm 0mm, clip, width=13cm]{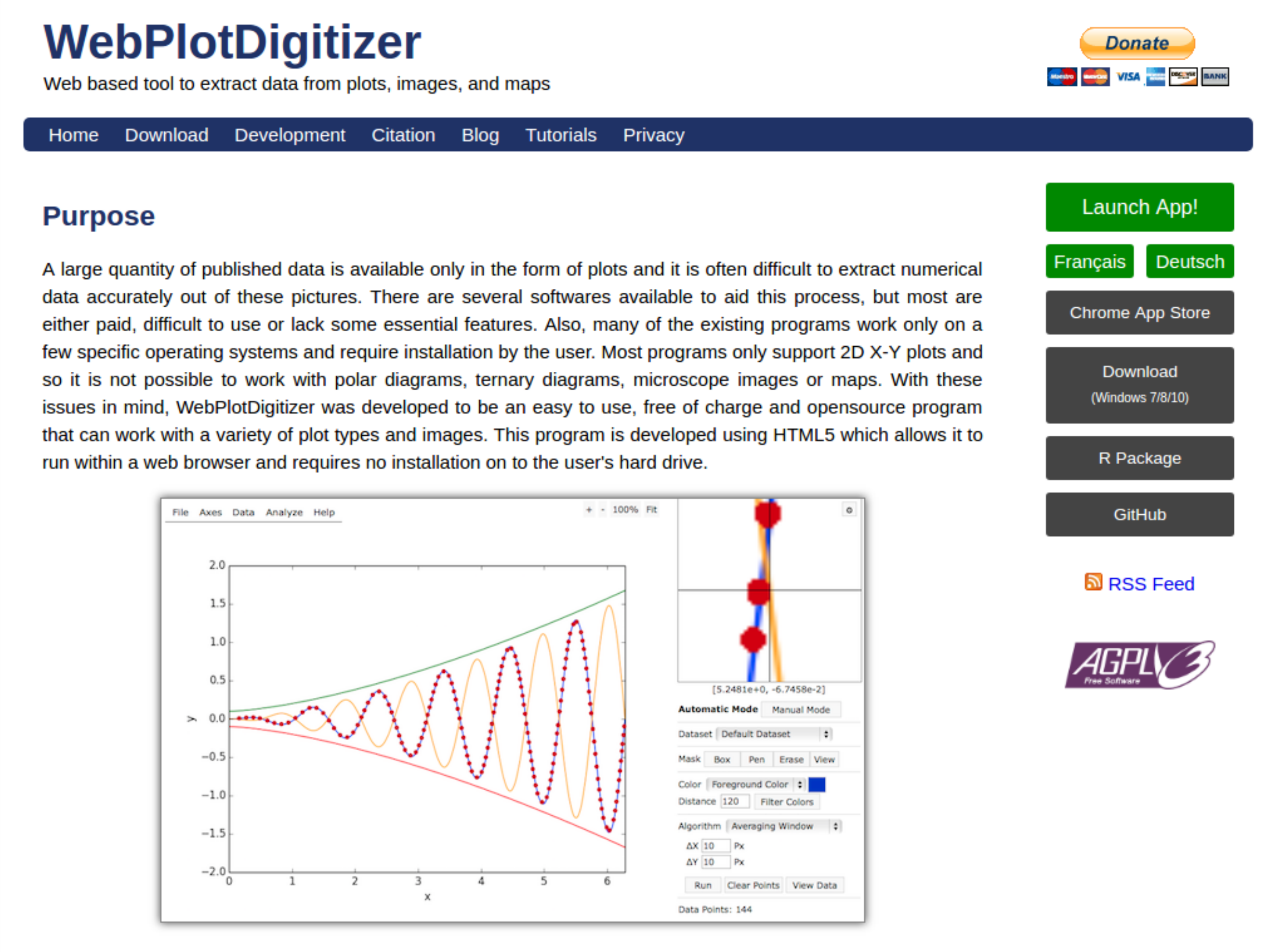}
  \caption{WebPlotDigitizer webpage.}
  \label{marin:fig1}
\end{figure}

WebPlotDigitizer has been continuously developed since its creation in 2011 by Ankit Rohatgi at the University 
of Notre Dame (Indiana, USA). This software uses affine transformations to map pixel location in the image to 
data points on a figure based on the calibration points provided by the user. Various image processing algorithms 
included with this software can be used to extract large number of data points from uploaded figures with great 
precision. WebPlotDigitizer is distributed under the GNU General Public License version 3, and the latest stable 
version of the software (v3.12, June 2017) can be used directly from the website \textcolor{blue}{http://arohatgi.info/WebPlotDigitizer}.
Fig.~\ref{marin:fig1} is a screenshot of the main web page. It shows the software launching button as well as all 
the recent developments of the code. A language selection is available (English, French and Deutsch). One can 
also download the code or donate for the code's sustainability.

\subsection{Extracting data from figures}

\begin{figure}[ht!]
 \centering
 \includegraphics[trim = 0mm 0mm 0mm 0mm, clip, width=16cm]{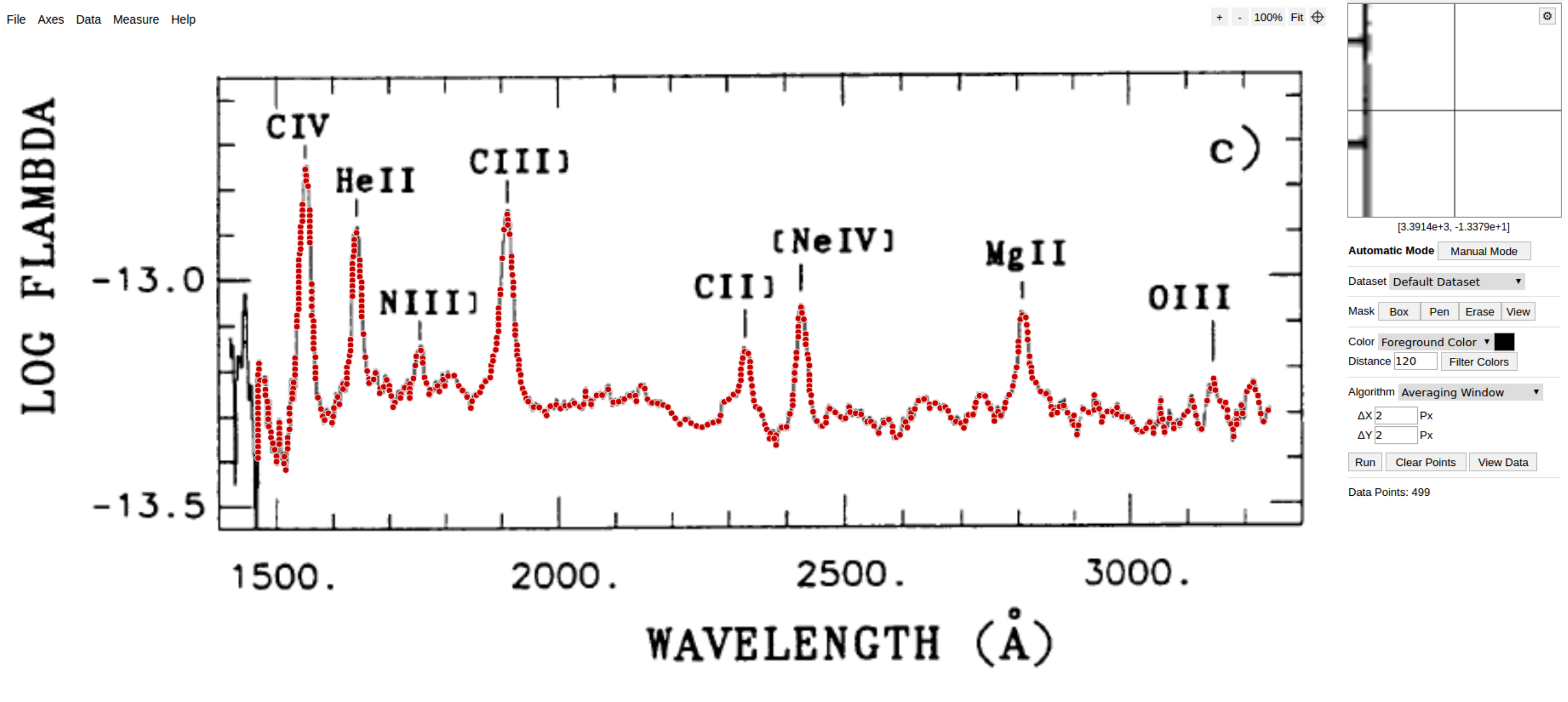}
  \caption{On-line interface of WebPlotDigitizer.
	   The red points are the calibrated data
	   sampled throughout the spectrum, with 
	   a resolution $\Delta$X and $\Delta$Y of 2
	   pixels (on-screen). The side panel shows 
	   the data acquisition controls. The spectrum 
	   is from \citet{Code1993}.}
  \label{marin:fig2}
\end{figure}

In order to extract a spectra from an archival paper, the first step is to create an image that isolates 
the spectrum, including the horizontal and vertical axes. The presence of text and legends won't interfere.
Once the image of the spectrum created (see, e.g., Fig.~\ref{marin:fig2}), it must be loaded by the program 
using File $>$ Load image. Different plot types can then be selected (regular XY plot, but also 2D bar plots, 
polar and ternary diagrams, maps with scale bar and images) to correctly map the image pixels to the 
corresponding data values in the image. The next step is to quantitatively calibrate the data using axes 
alignment. The software will ask the user to select a few known points on the axes (2 points on the X-axis 
and 2 more points on the Y-axis for a 2D spectrum) and enter the corresponding values. For better accuracy 
during the digitization process, it is better to pick points that are as far away from each other as possible. 

To acquire the data after plot axes calibration, several options are available. A tedious option is to manually 
select data points on the image (particularly helpful for low-resolution spectra with error bars). A zoomed
view, on the top-right corner of the screen, see Fig.~\ref{marin:fig2}, helps to be precise and reflects actual 
data coordinates corresponding to the mouse position on the image. The manual mode allows for point selection, 
adjustment or removal by manually clicking at the desired locations. An alternative, faster method, is to use 
the automatic mode. In the automatic mode, the user can set up and execute an extraction algorithm that can 
differentiate between the data points and the image background and identify several data points in a short time.
This is the method that was used to extract the total flux spectrum from NGC~1068 from \citet{Code1993} in 
Fig.~\ref{marin:fig2}.

Boxes of user-defined sizes and a virtual pen are used to define the region of interest where the spectrum resides. 
The user literally paints the spectrum, without spilling over the axes or legends, and the algorithms look for the 
foreground color specified for the data and ignore everything else. A background mode also allows the algorithms 
to include everything except the background color as potential data points, a very useful feature in the case 
of overlapping curves/spectra of different colors. If the user paints over the axes, then it is possible to erase 
the erroneous parts and paint again, without limits on the number of attempts. When the spectrum is fully colored,
the user simply has to press the Run button to start the auto-detection algorithm. After this is completed, 
the detected points appear in red over the image (see Fig.~\ref{marin:fig2}). If necessary, one can adjust 
the parameters of the extraction algorithm, mask or color settings and run the algorithm until satisfied. 

The last stage of the operation is to download the digitized values. By clicking the View Data button, a 
popup window appears and the values can be sorted by the variable or in order of the distance between the 
points (Nearest Neighbor). The final values can be copied and saved\footnote{A detailed user manual is 
available on the software webpage: http://arohatgi.info/WebPlotDigitizer/userManual.pdf, and additional 
examples are shown on YouTube videos.}.

\subsection{Application to ultraviolet spectropolarimetric data on NGC~1068}

\begin{figure}[ht!]
 \centering
 \includegraphics[trim = 0mm 0mm 0mm 0mm, clip, width=16cm]{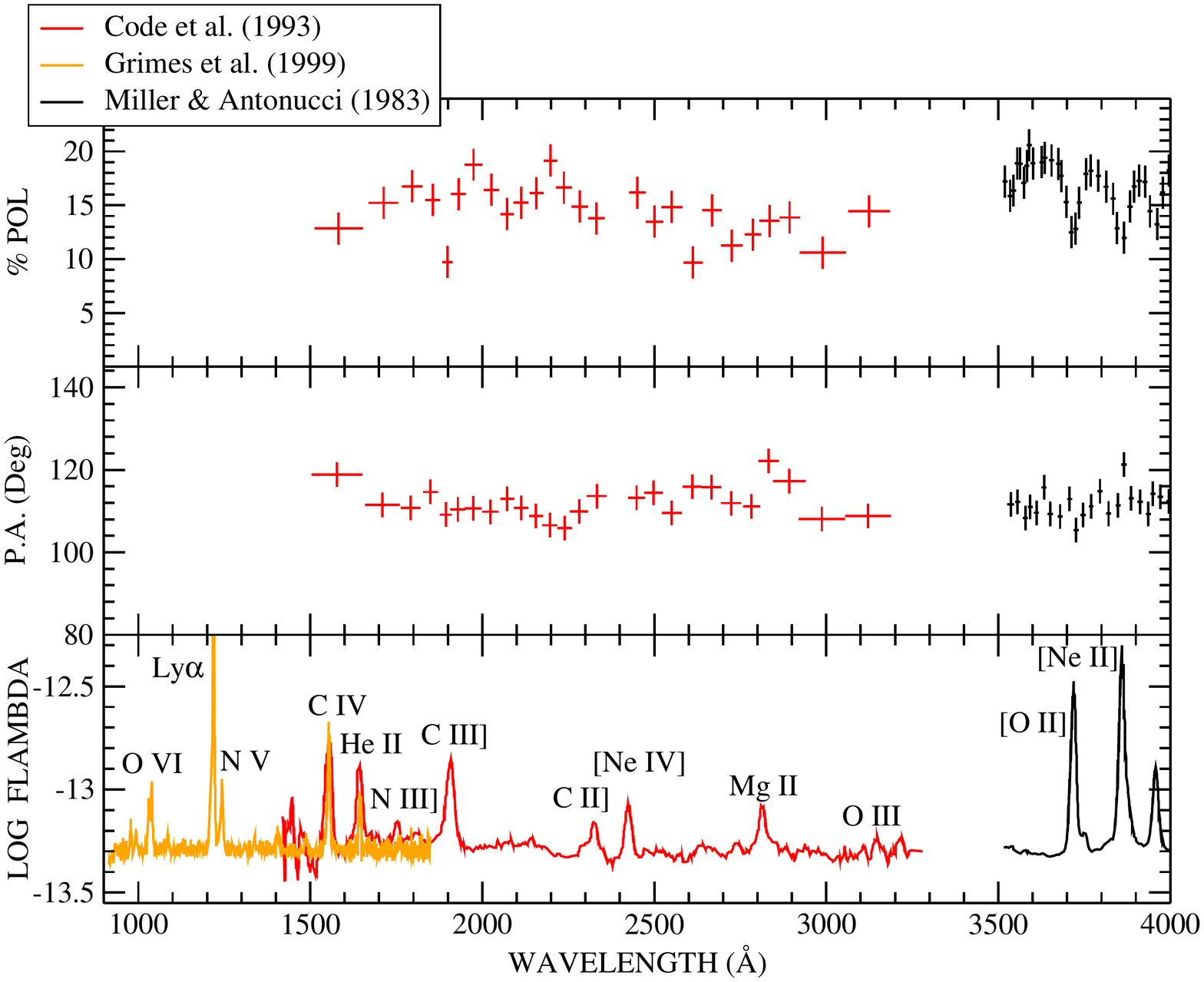}
  \caption{NGC~1068 ultraviolet polarization degree (top), 
	   polarization position angle (middle) and total 
	   flux (bottom) extracted from \citet{Miller1983}, 
	   \citet{Code1993} and \citet{Grimes1999} using
	   WebPlotDigitizer.}
  \label{marin:fig3}
\end{figure}

We used WebPlotDigitizer to extract the total flux spectra and polarization from old observations. Namely, 
we first extracted the ultraviolet flux spectrum (in log(F$_\lambda$)) from \citet{Grimes1999}. They observed 
the far-ultraviolet spectrum (912 -- 1840~\AA) of NGC~1068 using the Hopkins Ultraviolet Telescope 
during the 1995 March Astro-2 mission. The spectrum is represented in orange in Fig.~\ref{marin:fig3}. Due 
the pixel resolution on screen the data are not as smooth as in the publication (resolution degradation 
to 0.5~\AA) but all the emission lines and the absorption features are correctly duplicated, together with 
the proper fluxes levels. We then extracted the total flux spectrum and polarization degree and angle taken 
by the Lick Observatory 3~m Shane telescope mounted with a Cassegrain Pockels cell polarimeter \citep{Miller1983}.
The near-ultraviolet emission is shown in black in Fig.~\ref{marin:fig3}. Finally, in red, is the 
Hubble Space Telescope spectropolarimetric observation made by \citet{Code1993}. We merged the three total 
flux spectra by normalizing their narrow line emission fluxes and continuum emission.

The final composite polarization spectrum of NGC~1068 is shown in Fig.~\ref{marin:fig3}. This is a 
compilation of all the observed ultraviolet polarimetric data we know on this nearby, very bright type-2 
AGN whose optical polarized spectrum allowed in 1985 to understand the true nature of radio-quiet AGN \citep{Antonucci1985}.
The numerous gaps in wavelengths and the very low polarimetric data resolution for this source makes 
it a perfect target for new generations of space-based polarimeters. In this regards, POLLUX, a 
high-resolution ultraviolet spectropolarimeter for the LUVOIR space mission at NASA 
\citep{Stahl2015,Bolcar2016,France2016,Peterson2017} is the ideal instrument. Supported by CNES and 
developed by a European consortium, POLLUX is expected to provide very high resolution polarimetric 
observations of bright sources in the 900 -- 4000~\AA~band (desired wavelength coverage). The study of 
AGN at high polarimetric resolution will enable to probe the physics of outflowing winds with 
incredible precision. The launching site of the theoretical disk-born winds can be probed at 
the closest radii from the central supermassive black hole and the polarized emission line and continuum 
will be directly influence by the geometry and composition of the wind \citep{Marin2013,Marin2013a,Marin2013b}. 
Additionally, the composition of the dusty material, at radii larger than the physical limits imposed by 
dust sublimation, could be probed in a waveband where no terrestrial or space-born polarimetric data exists 
with sufficient wavelength coverage and resolution \citep{Hines2001}.

\section{Conclusions}
We have shown the ability of the WebPlotDigitizer software to precisely and efficiently digitize graphical 
spectra. Such an ability has potential implications for easy model-to-data comparison in many fields of astronomy. 
It can help access old data in record time without going through data reduction again and can be used, for example, 
to compile all existing spectra of a particular object. We applied WebPlotDigitizer to the case of NGC~1068, a bright 
type-2 Seyfert galaxy whose ultraviolet polarization spectrum lacks a proper coverage, both in wavelength and resolution.
By creating the first composite ultraviolet polarimetric spectrum of NGC~1068, we highlighted the need for a high-resolution
ultraviolet spectropolarimeter such as POLLUX.

\begin{acknowledgements}
F.M acknowledges funding through the CNES post-doctoral position grant 
``Probing the geometry and physics of active galactic nuclei
with ultraviolet and X-ray polarized radiative transfer''. 
\end{acknowledgements}

\bibliographystyle{aa}  % A&A bibliography style file (aa.bst)
\bibliography{marin} % your references in file: Yourfile.bib

\end{document}